\documentclass{LMCS}

\usepackage{latexsym} 
\usepackage{amssymb,amsmath}
\usepackage{epsfig}
\usepackage{enumerate,hyperref}

\def\eqalign#1{\null\,\vcenter{\openup\jot
  \ialign{\strut\hfil$\displaystyle{##}$&$\displaystyle{{}##}$\hfil
      \crcr#1\crcr}}\,}

\def\nat{{\mathbb N}}
\def\real{{\mathbb R}}

\newcommand{\pl}{\mbox{\tt pl}}

\newcommand{\bfx}{\mathbf{x}}

\newcommand{\bfq}{\mathbf{q}}
\newcommand{\bfzero}{\mathbf{0}}

\newcommand{\bfqstar}{\mathbf{q}^*}
\newcommand{\Call}{\mbox{\em Call}}
\newcommand{\Val}{\mbox{Val}}

\newcommand{\Dist}{\mbox{${\mathcal D}$}}

\newcommand{\play}{{\mbox{\em play}}}
\newcommand{\Return}{\mbox{\em Return}}
\newcommand{\lfp}{\mbox{\rm LFP}}

\theoremstyle{plain}
\newtheorem{theorem}[thm]{Theorem}

\newtheorem{example}{Example}
 
\newtheorem{corollary}[thm]{Corollary}
 
\newtheorem{lemma}[thm]{Lemma}
\newtheorem{proposition}[thm]{Proposition} 

\newcommand{\rfotwo}
{\mbox{\rm FO}^2[<]}
\newcommand{\tl}{\mathrm{TL}}            

\newcommand{\utl}
{\mbox{{\rm unary-}}\tl}

\def\doi{4 (4:7) 2008}
\lmcsheading%
{\doi}
{1--21}
{}
{}
{May\phantom{.}~24, 2007}
{Nov.~11, 2008}
{}   

\begin{document}

\title{Recursive Concurrent Stochastic Games\rsuper *}

\titlecomment{{\lsuper *}A preliminary version of this paper
  appeared in the 
\emph{Proceedings of the 33rd International Colloquium on
Automata, Languages and Programming ({ICALP}'06).}}

\author[K. Etessami]{Kousha Etessami\rsuper a}
\address{{\lsuper a}LFCS, School of Informatics, University of Edinburgh, UK}
\email{kousha@inf.ed.ac.uk}

\author[M. Yannakakis]{Mihalis Yannakakis\rsuper b}
\address{{\lsuper b}Department of Computer Science, Columbia University, USA}
\email{mihalis@cs.columbia.edu}


\begin{abstract}
We study Recursive Concurrent Stochastic Games (RCSGs),
extending our recent analysis of 
recursive simple stochastic games 
to a concurrent setting where the two
players choose moves simultaneously and independently at each state.
For multi-exit games, our earlier work already showed undecidability 
for basic questions like termination, thus we focus
on the important case of single-exit RCSGs (1-RCSGs).

We first characterize the value of a 1-RCSG termination game 
as the least fixed point solution of a system of 
nonlinear minimax functional equations, and use it to  
show PSPACE decidability for the quantitative termination problem.
We then give a strategy improvement
technique, which we use to show 
that player 1 (maximizer) has $\epsilon$-optimal
randomized Stackless \& Memoryless (r-SM) 
strategies for all $\epsilon > 0$, while player 2 (minimizer)
has optimal r-SM strategies.  
Thus, such games are r-SM-determined.
These results mirror and
generalize in a strong sense the
randomized memoryless determinacy results
for finite stochastic games, and   
extend the classic Hoffman-Karp 
strategy improvement approach from the finite
to an infinite state setting.
The proofs in our infinite-state setting are very different however,
relying on subtle analytic properties of certain power series
that arise from studying 1-RCSGs.

We show that our 
upper bounds, even for qualitative (probability 1) termination,
can not be improved, even to NP, without a
major breakthrough, by giving two reductions: first
a P-time reduction from the long-standing square-root sum problem
to the quantitative termination decision problem   
for {\em finite} concurrent stochastic games, 
and then a P-time reduction from the latter problem 
to the qualitative termination problem for 1-RCSGs.
\end{abstract}

\maketitle

\section{Introduction}
\noindent In recent work we have 
studied  Recursive Markov Decision Processes (RMDPs) 
and turn-based Recursive Simple Stochastic Games (RSSGs)
(\cite{EY05icalp,EY06stacs}), providing a number of strong
upper and lower bounds for their analysis.
These define infinite-state (perfect information) stochastic games
that extend Recursive Markov Chains (RMCs) (\cite{EY04a,EY05short}) 
with non-probabilistic actions controlled by  players.
Here we extend our study to
Recursive Concurrent Stochastic Games (RCSGs), where the two
players choose moves simultaneously and independently at each state,
unlike RSSGs where only one player can move at each state.  
RCSGs define a class of infinite-state zero-sum (imperfect information) 
stochastic games that can naturally model
probabilistic procedural programs and other systems involving both
recursive and probabilistic behavior, as well as concurrent interactions 
between the system and the environment.
Informally, all such recursive models consist of 
a finite collection
of finite state component models (of the same type)
that can call each other in a potentially recursive manner.
For RMDPs and RSSGs with multiple exits (terminating states), 
our earlier work already showed 
that basic questions such as almost sure termination
(i.e. does player 1 have a strategy that ensures termination 
with probability 1) are  already undecidable;
on the other hand, we 
gave strong upper bounds for the important  special case of 
{\em single-exit} 
RMDPs and RSSGs  (called 1-RMDPs and 1-RSSGs).

Our focus in this paper is thus on single-exit
Recursive Concurrent Stochastic Games (1-RCSGs for short).  
These models correspond to a concurrent game version of 
multi-type {\em Branching Processes} and {\em Stochastic Context-Free Grammars}, 
both of which are important
and extensively studied stochastic processes with many applications
including in population genetics, nuclear
chain reactions, 
computational biology,
and natural language processing
(see, e.g., \cite{Harris63,Jag75,KA02}
and other references in \cite{EY04a,EY05icalp}).
It is very natural to consider game 
extensions to these stochastic models.
Branching processes model the growth of a population of entities
of distinct types. In each generation each entity of a given type
gives rise, according to a probability distribution, 
to a multi-set of entities of distinct types. A branching process
can be mapped to a 1-exit Recursive Markov Chain (1-RMC) 
such that the probability of eventual extinction
of a species is equal to the probability of termination in the 1-RMC.
Modeling the process in a context where external agents can
influence the evolution to bias it towards extinction or towards
survival leads naturally to a game. A 1-RCSG
models the process where the evolution of some types
is affected by the concurrent actions of
external favorable and unfavorable agents (forces).

In \cite{EY05icalp}, we showed that
for the turned-based 1-RSSG termination game, 
where the goal of player 1 (respectively, player 2) is to maximize (resp. minimize)
the probability of termination starting at a given vertex (in
the empty calling context),
we can decide in 
PSPACE whether the value of the game is $\geq p$ for a given
probability $p$, and we can approximate this value (which can
be irrational) to within given
precision with the same complexity.  We also showed that both players have 
optimal {\em deterministic} 
{\em Stackless and
Memoryless} (SM) strategies in the 1-RSSG termination game;  
these are strategies that depend neither on the history of the
game nor on the call stack at the current state.
Thus from each vertex belonging to the player, such a strategy 
deterministically picks 
one of the outgoing transitions.

Already for finite-state concurrent stochastic 
games (CSGs), even under the simple termination
objective, 
the situation is rather
different.   Memoryless strategies do suffice for both players, 
but randomization of strategies is necessary,
meaning 
we can't hope for deterministic $\epsilon$-optimal strategies
for either player.    Moreover, player 1 (the maximizer) 
can only attain $\epsilon$-optimal strategies,
for $\epsilon > 0$,
whereas player 2 (the minimizer) does have optimal randomized
memoryless strategies (see, e.g., \cite{FiVr97,dAMa04}). 
Another important result for finite CSGs is the classic 
Hoffman-Karp \cite{HofKar66} strategy improvement method,
which provides, via simple local improvements, a sequence of 
randomized memoryless strategies which yield payoffs  
that converge to the value of the game.

Here we generalize all these results to 
the infinite-state setting of 1-RCSG termination games.
We first characterize values of the 1-RCSG termination game 
as the least fixed point solution of a system of nonlinear minimax 
functional equations.
We use this to show PSPACE
decidability for the {\em qualitative termination
problem} (is the value of the game $=1$?) and the 
{\em quantitative termination problem}
(is the value of the game $\geq r$ (or $\leq r$, etc.), 
for given rational $r$),
as well as PSPACE algorithms for approximating the termination
probabilities of 1-RCSGs to within a given number of bits of precision,
via results for the existential theory of reals.
(The simpler ``qualitative problem'' of deciding 
 whether the game value is $= 0$
 only depends on the transition structure of the 1-RCSG and
not on the specific probabilities.  For this problem
 we give a polynomial time algorithm.)

We then proceed to our technically most involved result, a strategy improvement
technique for 1-RCSG termination games.   We use this to show that in
these games player 1 (maximizer) has $\epsilon$-optimal
randomized-Stackless \& Memoryless (r-SM for short) 
strategies, whereas player 2 (minimizer)
has optimal r-SM strategies.  
Thus, such games are r-SM-determined.
These results mirror and
generalize in a very strong sense the 
randomized memoryless determinacy results known  
for finite stochastic games.
Our technique 
extends Hoffman-Karp's 
strategy improvement method for finite CSGs
to an infinite state setting.
However, the proofs in our infinite-state setting are  
very different.  We  
rely on subtle analytic properties of certain power series
that arise from studying 1-RCSGs.

Note that our 
PSPACE upper bounds for the quantitative termination
problem for 1-RCSGs can not be improved to NP without a major breakthrough,
since already for 1-RMCs we showed in \cite{EY04a} 
that the quantitative termination
problem is at least as hard as the square-root sum problem 
(see \cite{EY04a}).
In fact, here we show that even the 
{\em qualitative termination problem} for 1-RCSGs, 
where the problem is to decide whether the value of the game is exactly 1,
is already as hard as the square-root sum problem, 
and moreover, so
is the quantitative termination decision problem for {\em finite} CSGs.
We do this via two reductions:  we give a P-time reduction
from the square-root sum problem to the 
quantitative termination decision problem for {\em finite} CSGs,
and a P-time reduction from the quantitative finite CSG termination
problem  to the qualitative 1-RCSG termination problem.

It is known (\cite{Chatpr}) that for finite concurrent games,
probabilistic nodes do not add any power to these games, because the
stochastic nature of the games can in fact be simulated by concurrency
alone. The same is true for 1-RCSGs.  Specifically, given a finite CSG
(or 1-RCSG), $G$, there is a P-time reduction to a finite concurrent
game (or 1-RCG, respectively) $F(G)$, without any probabilistic
vertices, such that the value of the game $G$ is exactly the same as
the value of the game $F(G)$.  We will provide a proof of this in
Section \ref{sec:basics} for completeness.\medskip

\noindent{\bf Related work.}  
Stochastic games go back to Shapley \cite{Shapley53}, 
who considered finite concurrent stochastic games 
with (discounted) rewards. See, e.g., 
\cite{FiVr97} for a recent book on stochastic games.
Turn-based ``simple'' finite stochastic games were 
studied by Condon \cite{Condon92}.
As mentioned, we studied RMDPs and (turn-based) RSSGs 
and their quantitative and qualitative termination problems 
in \cite{EY05icalp,EY06stacs}. 
In \cite{EY06stacs} we showed that
the qualitative termination problem for both maximizing and
minimizing 1-RMDPs is in P,
and for 1-RSSGs is in NP$\cap$coNP.
Our earlier work \cite{EY04a,EY05short}  
developed theory and algorithms for Recursive Markov Chains (RMCs),
and \cite{EKM04,BKS05} have studied probabilistic 
Pushdown Systems which are essentially equivalent to RMCs.

Finite-state concurrent stochastic games have been
studied extensively in recent CS literature 
(see, e.g., \cite{CdeAHen06,dAMa04,dAHK98}).
In particular, the papers \cite{CMJ04} and \cite{CdeAHen06} have 
studied, for finite CSGs, the {\em approximate} reachability 
problem and
{\em approximate} parity game problem, respectively.
In those papers, it was claimed that these approximation 
problems are in NP$\cap$coNP.  Actually there was a minor
problem with the way the results on approximation were phrased
in \cite{CMJ04,CdeAHen06},
as pointed out in the conference version of this paper
\cite{EY06icalp}, but this is a 
relatively unimportant point compared to the flaw we
shall now discuss.
There is in fact a serious flaw in a key proof
of \cite{CMJ04}.  The flaw relates to the use of
a result from \cite{FiVr97} which shows 
that for discounted stochastic games the value function
is Lipschitz continuous with respect to the coefficients
that define the game as well as the discount $\beta$.  
Importantly, the Lipschitz constant in this result from
\cite{FiVr97} depends on the discount $\beta$ (it is inversely
proportional to $1-\beta$).
This fact was unfortunately overlooked in \cite{CMJ04} 
and, at a crucial point in their proofs, the Lipschitz constant was assumed to be a fixed constant 
that does not depend on $\beta$.  
This flaw unfortunately affects
several results in \cite{CMJ04}.  
It also affects the results of \cite{CdeAHen06}, since
the later paper uses the reachability results of \cite{CMJ04}. 
As a consequence of this error, the best upper bound which 
currently follows 
from the results in \cite{CMJ04,dAMa04,CdeAHen06}
is a PSPACE upper bound for the decision and approximation
problems for the value of finite-state concurrent stochastic reachability 
games as well as for finite-state concurrent stochastic parity games.
(See the erratum note for \cite{CMJ04} 
on K. Chatterjee's web page \cite{Chat-err07},
 as well as his Ph.D. thesis.) 
It is entirely plausible that these results can be repaired and that
approximating the value of finite-state concurrent reachability
games to within a given additive error $\epsilon > 0$ 
can in the future be shown to be in NP $\cap$ coNP,
but the flaw in the proof given in \cite{CMJ04} is fundamental
and does not appear to be easy to fix.

On the other hand, for the quantitative decision problem 
for finite CSGs (as opposed to the approximation problem), 
and even the qualitative decision problem 
for 1-RCSGs, the
situation is different.
We show here that the quantitative decision problem for finite
CSGs, as well as the qualitative decision problem for 1-RCSGs, are
both 
as hard as the square-root sum problem, for which containment even 
in NP is a long standing open problem.
Thus our PSPACE upper bounds here,  even
for the qualitative termination 
problem for 1-RCSGs, can not be improved to NP without a major
breakthrough.
Unlike
for 1-RCSGs,
the qualitative termination problem 
for finite CSGs
is known to be decidable in P-time (\cite{dAHK98}).
We note that in recent work Allender et. al. 
\cite{ABKM06} have shown that the square-root sum problem
is in (the 4th level of) the ``Counting Hierarchy'' CH,
which is inside PSPACE, but it remains 
a major open problem to bring this complexity down to NP.

The rest of the paper is organized as follows.
In Section 2 we present the RCSG model, define the problems
that we will study, and give some basic properties.
In Section 3 we give a system of equations that characterizes the
desired probabilities, and use them to show that the problems
are in PSPACE. In Section 4 we prove the existence of 
optimal randomized stackless and memoryless strategies, and we
present a strategy improvement method.
Finally in Section 5 we present reductions from the square root sum
problem to the quantitative termination problem for finite
CSGs, and from the latter to the qualitative problem for
Recursive CSGs.

\vspace{-0.16in}

\section{Basics}
\label{sec:basics}

\noindent We have two players, Player 1 and Player 2.
Let $\Gamma_1$ and $\Gamma_2$ be finite sets  
constituting the {\em move alphabet} of players 1 and 2, respectively.
Formally, a {\em Recursive Concurrent Stochastic Game (RCSG)} 
is a tuple $A = (A_1, \ldots ,A_k)$, where 
each {\em component \/}
$A_i = (N_i , B_i, Y_i, En_i, Ex_i, \pl_i,\delta_i)$  consists of:
\begin{enumerate}[(1)]

\item A finite set $N_i$ of {\em nodes\/}, with a distinguished subset $En_i$
of {\em entry} nodes and a (disjoint) subset $Ex_i$ of {\em exit} nodes. 

\item A finite set $B_i$ of {\em boxes\/}, 
and a mapping $Y_i: B_i  \mapsto \{ 1, \ldots,k \}$ 
that assigns to every box (the index of) a component. 
To each box $b \in B_i$, we associate a set of {\em call ports}, $\Call_b = 
\{ (b,en) \mid en \in En_{Y(b)} \}$, 
and a set of {\em return ports}, $\Return_b = \{ (b,ex) \mid ex \in Ex_{Y(b)}\}$.
Let $\Call^i = \cup_{b \in B_i} \Call_b$, 
$\Return^i = \cup_{b \in B_i} \Return_b$,
and let $Q_i =N_i \cup Call^i \cup Return^i$ be the set of all nodes, call ports and
return ports; we refer to these as the {\em vertices} of component $A_i$.

\item A mapping $\pl_i: Q_i \mapsto \{0,\play\}$ that assigns 
to every vertex $u$ a type describing how the next transition is chosen:
if $\pl_i(u)$ $=$ $0$ it is chosen probabilistically and if $\pl_i(u)$ $=$ $play$ it
is determined by moves of the two players.
Vertices $u \in (Ex_i \cup \Call^i)$ have no outgoing transitions;
for them we let $\pl_i(u) = 0$.

\item A transition relation 
$\delta_i \subseteq (Q_i \times (\real \cup (\Gamma_1 \times  \Gamma_2)) \times Q_i)$, 
where for each tuple $(u,x,v) \in \delta_i$,
the source $u \in (N_i \setminus Ex_i) \cup \Return^i$, the destination
$v \in (N_i \setminus En_i) \cup \Call^i$, where if $\pl(u) =0$ then $x$ is 
a real number $p_{u,v} \in [0,1]$ (the transition probability), 
and if $\pl(u) = \play$ then $x=(\gamma_1,\gamma_2) \in \Gamma_1 \times \Gamma_2$.
We assume that each vertex $u \in Q_i$ has associated with it a set 
$\Gamma^u_1 \subseteq \Gamma_1$ and a set $\Gamma^u_2 \subseteq \Gamma_2$,
which constitute player 1 and  2's {\em legal moves} at vertex $u$.
Thus, if $(u,x,v) \in \delta_i$ and $x=(\gamma_1,\gamma_2)$  then 
$(\gamma_1,\gamma_2) \in \Gamma^u_1 \times \Gamma^u_2$.    
Additionally,  for each vertex $u$ and each $x \in \Gamma^u_1 \times \Gamma^u_2$, we
assume there is exactly one transition of the form $(u,x,v)$ in $\delta_i$. 
Furthermore they must satisfy the consistency property:
for every $u \in \pl^{-1}(0)$, 
$\sum_{\{v' \mid (u,p_{u,v'},v') \in \delta_i\}} p_{u,v'} = 1$,
unless $u$ is a call port or exit node, neither of which have 
outgoing transitions, in which case by default
$\sum_{v'} p_{u,v'} = 0$.
\end{enumerate}

\begin{figure*}[t]
\begin{center}
\scalebox{0.55}{\input{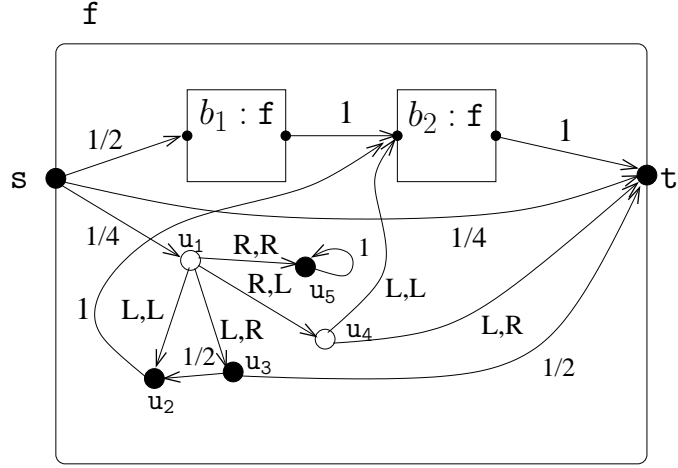}}
\caption{Example (1-exit) RCSG}
\label{fig:example_1rcsg}
\end{center}
\end{figure*}

\noindent We use the symbols ($N, B, Q, \delta,$ etc.) without a subscript, 
to denote the union over all components. 
Thus, eg. $N =\cup^k_{i=1} N_i$ is the set of all nodes of $A$,
$\delta=\cup^k_{i=1}\delta_i$ the set of all transitions,
$Q  =\cup^k_{i=1} Q_i$ the set of all vertices, etc.
The set $Q$ of vertices is partitioned into the sets
$Q_{play} = \pl^{-1}(\play)$ and $Q_{prob} = \pl^{-1}(0)$
of play and probabilistic vertices respectively.

For computational purposes we assume that the transition probabilities 
$p_{u,v}$ are rational, given in the input as the ratio of two
integers written in binary. The {\em size} of a RCSG is the
space (in number of bits) needed to specify it fully,
i.e., the nodes, boxes, and transitions of all components,
including the probabilities of all the transitions.

\begin{example}
{\rm
An example picture of a (1-exit) RCSG is depicted in Figure \ref{fig:example_1rcsg}. 
This RCSG has one component, $f$, which has 
nodes $\{{s,t,u_1,u_2,u_3,u_4,u_5}\}$.
It 
has one entry node, ${s}$, and one exit node, ${t}$.
It also has two boxes, $\{b_1, b_2\}$, both of which map to the only component, ${f}$.
All nodes in this RCSG are probabilistic (black nodes) except for nodes 
${u_1}$ and ${u_4}$ 
which are player nodes (white nodes).  The move alphabet for both players is $\{L,R\}$
(for, say, ``left'' and ``right'').  At node ${u_1}$ both players have both moves enabled.
At node ${u_4}$, player 1 has only $L$ enabled, and player 2 has both $L$ and $R$
enabled.  \qed
}

\label{example:rcsg}
\end{example}

An RCSG $A$ defines a global denumerable stochastic game 
$M_A = (V,\Delta,\pl)$ as follows.
The global {\em states\/} $V \subseteq B^* \times Q$  of $M_A$
are pairs of the form $\langle \beta, u \rangle$, where $\beta \in B^*$
is a (possibly empty) sequence of boxes and $u \in Q$ is a {\em vertex} of $A$.
More precisely, the states $V\subseteq B^* \times Q$ and transitions $\Delta$ are defined inductively as follows:
\begin{enumerate}[(1)]

\item $\langle \epsilon, u \rangle \in V$, for $u \in Q$ 
($\epsilon$ denotes the empty string.)  

\item If $\langle \beta, u \rangle \in V$ 
and $(u,x,v) \in \delta$, then
$\langle \beta,v \rangle \in V$ and 
$(\langle \beta, u \rangle, x,\langle \beta,v \rangle) \in \Delta$.

\item If $\langle \beta , (b,en) \rangle \in V$, with $(b,en) \in \Call_b$, 
then $\langle \beta b, en\rangle \in V$ and
$(\langle \beta , (b,en) \rangle, 1, \langle \beta b, en\rangle ) \in \Delta$. 

\item If $\langle \beta b, ex \rangle \in V$, and $(b,ex) \in \Return_b$,
then $\langle \beta,(b,ex) \rangle \in V$ and
$(\langle \beta b, ex \rangle, 1, \langle \beta,(b,ex) \rangle)\!\in \Delta$. 
\end{enumerate}
 
Item 1 corresponds to the possible initial states,
item 2 corresponds to control staying within a
component, item 3 is when a new component is entered
via a box, item 4 is when control exits a box
and returns to the calling component.
The mapping $\pl: V \mapsto \{0,\play\}$ is given
by $\pl(\langle \beta, u \rangle) = \pl(u)$.
The set of vertices $V$ is partitioned into 
$V_{prob}$, $V_\play$,  
where $V_{prob} = \pl^{-1}(0)$ and $V_\play = \pl^{-1}(\play)$.

We consider $M_A$ with various {\em initial states} of the form 
$\langle \epsilon, u \rangle$,  denoting this by $M^u_A$.
Some states of $M_A$ are {\em terminating states} and have no outgoing transitions.
These are states  $\langle \epsilon,ex\rangle$, where $ex$ is an exit node.  
If we wish to view $M_A$ as a non-terminating CSG, we can consider the
terminating states as absorbing states of $M_A$, with a self-loop of probability 1.

An RCSG where $| \Gamma_2 | = 1$
(i.e., where player 2 has only one action) is called a maximizing {\em Recursive Markov Decision Process} (RMDP); likewise, when $|\Gamma_1 | = 1$ 
the RCSG is a minimizing RMDP.
An RSSG where $|\Gamma_1| = | \Gamma_2 | = 1$ 
is essentially a {\em Recursive Markov Chain}
(\cite{EY04a,EY05short}).

Our goal is to answer termination 
questions for RCSGs of the form: {\em ``Does player 1 have a strategy to 
force the game to terminate (i.e., reach node $\langle \epsilon, ex \rangle$), 
starting at $\langle \epsilon, u \rangle$, with probability $\geq p$,
regardless of how player 2 plays?''}.

First, some definitions: a {\em strategy} $\sigma$ for
player $i$, $i \in \{1,2\}$,  is a function $\sigma: V^*V_\play \mapsto \Dist(\Gamma_i)$,
where $\Dist(\Gamma_i)$ denotes the set of probability distributions 
on the finite set of moves $\Gamma_i$.
In other words,  given a history $ws \in V^*V_\play$, and a strategy $\sigma$ for, say, player 1, 
$\sigma(ws)(\gamma)$ defines the probability with which player 1 will play move $\gamma$.
Moreover, we require that the function $\sigma$ has the property that for any 
global state $s = \langle \beta,u \rangle$,
with $\pl(u) = \play$, 
$\sigma(ws) \in \Dist(\Gamma^u_i)$.
In other words, the distribution has support only over eligible moves at vertex $u$.

Let $\Psi_i$ denote the set of all strategies for player $i$.
Given a history $ws \in V^*V_\play$ of play so far,
and given a 
strategy $\sigma \in \Psi_1$ for player $1$, and a strategy
$\tau \in \Psi_2$ 
for player $2$, 
the strategies determine a distribution on the next 
move of play to a new global state, namely, the transition $(s,(\gamma_1,\gamma_2),s') \in \Delta$
has probability $\sigma(ws)(\gamma_1) * \tau(ws)(\gamma_2)$.
This way, given a start node $u$, a
strategy $\sigma \in \Psi_1$, and a strategy
$\tau \in \Psi_2$, 
 we define a new Markov chain  (with initial state $u$) 
$M^{u,\sigma,\tau}_A = ({\mathcal S},\Delta')$.  
The states ${\mathcal S} \subseteq \langle \epsilon, u \rangle V^*$ of 
$M^{u,\sigma,\tau}_A$ are non-empty sequences of states of $M_A$,
which must begin with  
$\langle \epsilon, u \rangle$.   
Inductively, if $ws \in {\mathcal S}$, then:
(0) if $s \in V_{prob}$ and $(s,p_{s,s'},s') \in \Delta$ then
$wss' \in {\mathcal S}$ and $(ws,p_{s,s'},wss') \in \Delta'$;
(1) if $s \in V_\play$, 
where $(s,(\gamma_1,\gamma_2),s') \in \Delta$,
then  if $\sigma(ws)(\gamma_1) >0$ and $\tau(ws)(\gamma_2) >0$
then
$wss' \in {\mathcal S}$ and 
$(ws,p,wss') \in \Delta'$, where $p= \sigma(ws)(\gamma_1) * \tau(ws)(\gamma_2)$.

Given initial vertex $u$, and final exit $ex$ in the same component,
and given strategies $\sigma \in \Psi_1$ and 
$\tau \in \Psi_2$,  for $k \geq 0$, 
let $q^{k,\sigma,\tau}_{(u,ex)}$ 
be the probability that, in $M^{u,\sigma,\tau}_A$, starting at initial
state $\langle \epsilon,u \rangle$, we will reach a state 
$w \langle \epsilon,ex \rangle$ in at most $k$ ``steps'' 
(i.e., where $|w| \leq k$).
Let $q^{*,\sigma,\tau}_{(u,ex)}
= 
\lim_{k \rightarrow \infty} q^{k,\sigma,\tau}_{(u,ex)}$
be the probability of ever terminating at $ex$, i.e.,
reaching $\langle \epsilon,ex \rangle$.  (Note, the limit exists:
it is a monotonically non-decreasing sequence bounded by $1$).
Let 
$q^k_{(u,ex)} = \sup_{\sigma \in \Psi_1} 
\inf_{\tau \in \Psi_2} q^{k,\sigma,\tau}_{(u,ex)}$
and let $q^*_{(u,ex)} = \sup_{\sigma \in \Psi_1} 
\inf_{\tau \in \Psi_2} q^{*,\sigma,\tau}_{(u,ex)}$.
For a strategy $\sigma \in \Psi_1$,  let
$q^{k,\sigma}_{(u,ex)} =  \inf_{\tau \in \Psi_2} q^{k,\sigma,\tau}_{(u,ex)}$,
and let $q^{*,\sigma}_{(u,ex)} = 
\inf_{\tau \in \Psi_2} q^{*,\sigma,\tau}_{(u,ex)}$.
Lastly, given a strategy $\tau \in \Psi_2$, let
$q^{k,\cdot,\tau}_{(u,ex)} = \sup_{\sigma \in \Psi_1} 
q^{k,\sigma,\tau}_{(u,ex)}$,   and let 
$q^{*,\cdot,\tau}_{(u,ex)} = \sup_{\sigma \in \Psi_1} 
q^{*,\sigma,\tau}_{(u,ex)}$.

From, general determinacy results (e.g., ``Blackwell determinacy'' \cite{Martin98}  
which applies to
all Borel two-player zero-sum stochastic games with countable state spaces; 
see also \cite{MaitraSudderth98}) 
it follows that the games  $M_A$ are {\em determined}, meaning:\\
$\sup_{\sigma \in \Psi_1} 
\inf_{\tau \in \Psi_2} q^{*,\sigma,\tau}_{(u,ex)}
= \inf_{\tau \in \Psi_2}\sup_{\sigma \in \Psi_1}  
q^{*,\sigma,\tau}_{(u,ex)}$.

We call a strategy $\sigma$ for either player a (randomized) 
{\em Stackless and Memoryless }({\em r-SM}) strategy if it neither 
depends on the history of the game,
nor on the current call stack.  In other words, a r-SM strategy 
$\sigma$ for player $i$ is given 
by a function $\sigma: Q_\play \mapsto \mathcal{D}(\Gamma_i)$, 
which maps each play vertex $u$ of the RCSG to  a probability 
distribution $\sigma(u) \in 
\mathcal{D}(\Gamma^u_i)$ on the moves available to player $i$ at vertex $u$.

We are interested in the following computational problems.
\begin{enumerate}[(1)]
\item The {\em qualitative} termination problem: Is $q^*_{(u,ex)} = 1$?
\item The {\em quantitative} termination (decision) problem: \\
given 
$r \in [0,1]$, is $q^*_{(u,ex)}   \geq r$?   Is $q^*_{(u,ex)} \leq r$?\\
The {\em approximate} version: approximate 
$q^*_{(u,ex)}$ to within desired precision.
\end{enumerate}

\noindent Obviously, the qualitative termination problem is a special case
of the quantitative problem, setting $r=1$. 
As mentioned, for multi-exit RCSGs these are
all undecidable. Thus we focus on {\em single-exit} RCSGs ({\em 1-RCSGs}),
where every component has one exit.
Since for 1-RCSGs it is always clear which exit  
we wish to terminate at starting at vertex $u$
(there is only one exit in $u$'s component), 
we abbreviate $q^*_{(u,ex)}, q^{*,\sigma}_{(u,ex)}$, etc., 
as $q^*_u, q^{*,\sigma}_u$, etc., and we likewise abbreviate
other subscripts.

A different ``qualitative'' problem is to ask whether $q^*_u = 0$?
As we will show in Proposition \ref{prop:zeros}, this
is an easy problem: deciding whether $q^*_u = 0$ for a vertex $u$ in 
a 1-RCSG can be done in polynomial time, and only depends on the 
transition structure
of the 1-RCSG, not on the specific probabilities.

As mentioned in the introduction, 
it is known that for concurrent stochastic games,
probabilistic nodes do not add any power, and
can in effect be ``simulated'' by concurrent
nodes alone (this fact
was communicated to us by K. Chatterjee  
\cite{Chatpr}).
 The same fact is true for 1-RCSGs.
Specifically, the following holds:

\begin{proposition} 
There is a P-time reduction $F$, which,
given
a finite CSG (or a 1-RCSG), $G$, computes 
a finite concurrent game (or 1-RCG, respectively) 
$F(G)$, without any
probabilistic vertices, such that the value of the game $G$ is exactly 
the same as the value of the game $F(G)$.
\end{proposition}

\begin{proof}
First, suppose for now that in $G$ all probabilistic
transitions have probability $1/2$. 
In other words, suppose that for a probabilistic vertex $s \in \pl^{-1}(0)$
(which is not an exit or a call port)
in an 1-RCSG,
we have  two transitions $(s,1/2,t) \in \delta$ and $(s,1/2,t') \in \delta$.
In the new game $F(G)$,  change $s$ to a play vertex, i.e., 
let $\pl(s) = \play$,
and let $\Gamma^s_1 = \Gamma^s_2 = \{a,b\}$,
and  replace the  probabilistic transitions
out of $s$ with the following 4 transitions:
$(s,(a,b),t)$, $(s,(b,a),t)$, $(s, (a,a),t')$ and
$(s,(b,b),t')$. 
Do this for all probabilistic vertices in $G$, thus obtaining $F(G)$
which contains no probabilistic vertices.

Now, consider any strategy $\sigma$ for player 1 in the
original game $G$, and a strategy $\sigma '$ in the new game $F(G)$
that is consistent with $\sigma$, i.e. for each history ending at
an original play vertex $\sigma '$ has the same distribution as $\sigma$
(and for the other histories ending at probabilistic vertices
it has an arbitrary distribution).  
For any strategy $\tau$ for player 2 in the game $G$,
consider the strategy, $F(\tau)$, for 
player 2 in $F(G)$, which is defined as follows:
whenever the play reaches a probabilistic vertex $s$ of $G$ 
(in any context and with any history)  
$F(\tau)$ plays $a$ and $b$ with 1/2 probability each.
At all non-probabilistic vertices of $G$,  $F(\tau)$ plays
exactly as $\tau$ (and it may use the history, etc.).
This way, no matter what player 1 does, whenever play reaches
the vertex $s$ (in any context) the
play will move from $s$ to $t$ and to $t'$  with probability $1/2$ 
each.
Thus for any vertex $u$, the value $q^{\star,\sigma,\tau}_u$ in the game $G$
is the same as the value $q^{*,\sigma',F(\tau)}_u$ in the game $F(G)$.
So the optimal payoff value for player 1 in the game 
starting at any vertex $u$ is not greater in $F(G)$ than in $G$.  
A completely symmetric argument shows that for player 2 
the optimal payoff value starting at $u$ 
is not greater in $F(G)$ than in $G$.
Thus, the value of the game starting at $u$ is the same in both games. 

We can now generalize this to arbitrary rational probabilities on transitions,
instead of just probability $1/2$, by using a basic trick to encode
arbitrary finite probability distributions using a 
polynomial-sized finite Markov
chain  all of whose transitions have probability $1/2$.
Namely, suppose $u$ goes to $v_1$ with probability $p/q$ and to 
$v_2$ with
probability $1-p/q$, where $p$,$q$ are integers with $k$ bits
(we can write both as $k$-bit numbers, by adding
leading 0's to $p$ if necessary so that it has length exactly $k$,
same as $q$).
Flip (at most) $k$ coins.
View this as generating a $k$ bit binary number.
If the number that comes out is $< p$ (i.e. $0,\ldots,p-1$), 
then go to $v_1$,
if between $p$ and $q$ (i.e., $p,\ldots, q-1$) then go to $v_2$, 
if $\geq q$ go back to the start, $u$.
A naive way to do this would require exponentially many states in $k$.
But we only need at most $2k$ states to encode this 
if we don't necessarily flip all $k$ coins
but rather do the transition to $v_1, v_2$ or $u$,
as soon as the outcome is clear from the
coin flips.
That is, if the sequence $\alpha$ formed by the initial
sequence of coin flips so far differs from both the
prefixes $p', q'$ of $p$ and $q$ of the same length, 
then we do the transition: if $\alpha < p'$ transition to $v_1$,
if $p' < \alpha < q'$ transition to $v_2$, and if
$\alpha > q'$ then transition to $u$.
Thus, we only need to remember the number $j$ of coins flipped so far,
and if $j$ is greater than the length of the common prefix of $p$ and $q$
then we need to remember also whether the coin flips 
so far agree with $p$ or with $q$.

Clearly, a simple generalization of this argument works for generating
arbitrary finite rational probability distributions 
$p_1/q, p_2/q, \ldots, p_r/q$,  such that $\sum^r_{i=1} (p_i/q)  = 1$.
If $q$ is a $k$-bit integer, then the number of new states needed is at most $rk$,
i.e. linear in the encoding length of the rationals $p_1/q,\ldots, p_r/q$.
\end{proof}

\section{Nonlinear minimax equations for 1-RCSGs}

\noindent In (\cite{EY05icalp}) we defined a monotone system $S_A$ 
of nonlinear min-{\small \& }-max equations for 1-RSSGs
(i.e. the case of simple games),
and showed that its {\em least fixed point} solution 
yields the desired probabilities $q^*_{u}$.
Here we generalize these to nonlinear minimax systems for concurrent games,
1-RCSGs.
Let us use a variable $x_{u}$ for each unknown $q^*_{u}$, and let 
$x$ be the vector of all $x_u$ , $u\in Q$.
The system $S_A$ has one equation of the form $x_u= P_u(\mathbf{x})$
for each vertex $u$.
Suppose that $u$ is in component $A_i$ with (unique) exit $ex$.
There are 4 cases based on the ``{\em Type}'' of $u$.
\begin{enumerate}[(1)]
\item $u \in Type_1$:  $u = ex$. In this case: $ x_{u} = 1 $.
\item $u \in Type_{rand}$: $ \pl(u) = 0$ and 
$u \in (N_i \setminus \{ ex \})\cup \Return^i$. Then the equation is
$x_{u} =  \sum_{\{v \mid (u,p_{u,v},v) \in \delta\}} p_{u,v}x_{v}.$
(If $u$ has no outgoing transitions, this equation is by definition $x_{u}=0$.)

\item $u \in Type_{call}$: $u = (b,en)$ is a call port. 
The equation is $  x_{(b,en)} =  
x_{en} \cdot x_{(b,ex')} $,
where $ex' \in Ex_{Y(b)}$ is the unique exit of $A_{Y(b)}$.

\item $u \in Type_{\play}$. Then the equation is $x_{u} =  \Val(A_u(x))$,
where the right-hand side is defined as follows.
Given a value vector $x$, and a play 
vertex $u$, consider the
zero-sum matrix game given by matrix $A_u(x)$, 
whose rows are indexed by player 1's moves $\Gamma^u_1$
from node $u$, and whose columns are indexed by player 2's moves $\Gamma^u_2$.
The payoff to player 1 under the
pair of deterministic moves $\gamma_1 \in \Gamma^u_1$, and $\gamma_2 \in \Gamma^u_2$, is given by
$(A_u(x))_{\gamma_1,\gamma_2} := x_v$, where $(u,(\gamma_1,\gamma_2),v) \in \delta$.
Let $\Val(A_u(x))$ be the value of this zero-sum matrix game.
By von Neumann's minimax theorem, the value and optimal mixed strategies exist,
and they can be obtained by solving
a Linear Program with coefficients given by the $x_i$'s.
\end{enumerate}

\noindent In vector notation, we denote the system $S_A$ by $x = P(x)$.
Given 1-exit RCSG $A$, we can easily construct this system.
Note that the operator $P: \real^n_{\geq 0} \mapsto \real^n_{\geq 0}$  is {\em monotone}: for $x, y \in \real^n_{\geq 0}$, if $x \leq y$ then $P(x) \leq P(y)$.
This follows because for two  game matrices $A$ and $B$
of the same dimensions,
if $A \leq B$  (i.e., $A_{i,j} \leq B_{i,j}$ for all $i$ and $j$), then
$\Val(A) \leq \Val(B)$. 
Note that by definition of $A_u(x)$, for $x \leq y$, $A_u(x) \leq A_u(y)$.

\begin{exa}
We now construct the system of nonlinear minimax functional equations, $x=P(x)$,
associated with the 1-RCSG we encountered in Figure \ref{fig:example_1rcsg}
(see Example \ref{example:rcsg}).
We shall need one variable for every {\em vertex} of that 1-RCSG, to represent
the value of the termination game starting at that vertex, and we will
need one equation for each such variable.
Thus, the variables we need are $x_{s}, x_t, x_{u_1}, \ldots, x_{u_5}$, 
$x_{(b_1,s)}, x_{(b_1,t)}$,  $x_{(b_2,s)}$, $x_{(b_2,t)}$.
The equations are as follows:
\begin{equation*}\label{example:equations}
\eqalign{
x_t       & =  1\cr
x_s       & =  (1/2)x_{(b_1,s)} + (1/4) x_t + (1/4) x_{u_1}\cr
x_{u_5}    & =  x_{u_5}\cr
x_{u_2}    & =  x_{(b_2,s)}\cr
x_{u_3}    & =  (1/2) x_{u_2} + (1/2) x_t\cr
x_{(b_1,s)} & =   x_{s} x_{(b_1,t)}\cr
}
\qquad
\eqalign{
x_{(b_1,t)} & =  x_{(b_2,s)}\cr
x_{(b_2,s)} & =  x_s x_{(b_2,t)}\cr
x_{(b_2,t)} & =  x_t\cr
x_{u_1}    & =   \Val \left( \left[  \begin{array}{cc} 
                                x_{u_2} & x_{u_3}\\
                                x_{u_4} & x_{u_5}
                                \end{array}   \right] \right)\cr
x_{u_4}    & =  \Val \left(  \left[ \begin{array}{cc}
                               x_{(b_2,s)} & x_t
                              \end{array} \right] \right)
}
\end{equation*}
\end{exa}

\noindent We now identify a particular solution to $x = P(x)$, called
the {\em Least Fixed Point} ($\lfp$) solution, which gives precisely
the termination game values.  Define $P^1(x) = P(x)$, and define
$P^k(x) = P(P^{k-1}(x))$, for $k > 1$.  Let $q^* \in \real^n$ denote
the $n$-vector $q^*_{u}, u \in Q$ (using the same indexing as used for
$x$).  For $k \geq 0$, let $q^k$ denote, similarly, the $n$-vector
$q^k_{u}, u \in Q$.

\begin{theorem} 
\label{lfp-char-thm}
Let $x = P(x)$ be the system $S_A$ associated with 1-RCSG $A$. 
Then $q^* = P( q^*)$, and
for all $q' \in \real^n_{\geq 0}$, if 
$q' = P(q')$, then 
$q^* \leq { q'}$
(i.e., $q^*$ is the {\em Least Fixed Point}, 
of $P: \real^n_{\geq 0} \mapsto \real^n_{\geq 0}$).
Moreover, $\lim_{k \rightarrow \infty} P^k(\bfzero) \uparrow q^*$,
i.e., the ``value iteration'' sequence $P^k(\bfzero)$ converges monotonically
to the LFP, $q^*$. 
\end{theorem}
\begin{proof}
We first prove that $q^* = P(q^*)$.
Suppose $q^* \neq P(q^*)$.   
The equations for vertices $u$ of types $Type_{1},Type_{rand}$, and $Type_{call}$ can be used to 
define precisely 
the values $q_u^*$ in
terms of other values  $q^*_v$.
Thus, the only possibility is that $q^*_{u} \neq P_u(q^*)$
for some vertex $u$ of $Type_{\play}$.
In other words, $q^*_u \neq \Val(A_u(q^*))$.

Suppose  $q^*_u < \Val(A_u(q^*))$. To see that this can't happen,
we construct a strategy $\sigma$ for player 1 that achieves better. 
At node $u$, let player 1's strategy $\sigma$ play in one step its optimal
randomized minimax strategy in the game $A_u(q^*)$ (which exists according to the minimax theorem).
Choose $\epsilon >0$  such that $\epsilon < \Val(A_u(q^*))- q^*_u$.
After the first step, at any vertex $v$ player 1's strategy
$\sigma$ will play in such a way
that achieves a value $\geq q^*_v - \epsilon$  (i.e, an $\epsilon$-optimal 
strategy in the rest of the game, which must exist because the game is
determined).  
Let ${\mathbf \varepsilon}$ be an $n$-vector every
entry of which is $\epsilon$.  Now, the matrix game $A_u(q^*-{\mathbf \varepsilon})$ is just an additive translation of the matrix game $A_u(q^*)$, and thus it
has
precisely the same $\epsilon$-optimal strategies as the matrix game $A_u(q^*)$, and moreover 
$\Val(A_u(q^*-{\mathbf \varepsilon})) = \Val(A_u(q^*)) - \epsilon$.
Thus, by playing strategy $\sigma$, player 1 guarantees a value which is 
$\geq \Val(A_u(q^*-{\mathbf \varepsilon})) = \Val(A_u(q^*)) - \epsilon > q^*_u$,
which is a contradiction.  Thus $q^*_u \geq \Val(A_u(q^*))$.

A completely analogous argument works for player 2, 
and shows that $q^*_u \leq \Val(A_u(q^*))$.
Thus $q^*_u = \Val(A_u(q^*))$, and hence $q^* = P(q^*)$.

Next, we prove that if $q'$ is any vector such that $q' = P(q')$, then
$q^* \leq q'$.
Let $\tau'$ be the randomized stackless and memoryless strategy for player 2
that always picks, at any state $\langle \beta, u \rangle$, 
for play vertex $u \in Q_\play$, a mixed 1-step strategy which
is an optimal strategy in the matrix game $A_u(q')$.
(Again, the existence of such a strategy is guaranteed by the minimax theorem.)

\begin{lemma}
For all strategies $\sigma \in \Psi_1$ of player 1, 
and for all $k \geq 0$,  $q^{k,\sigma,\tau'} \leq q'$.
\end{lemma}
\proof
By induction.
The base case $q^{0,\sigma,\tau'} \leq q'$ is trivial.
\begin{enumerate}[(1)]
\item $Type_1$.  If $u = ex$ is an exit, then for all $k\geq 0$, 
clearly $q^{k,\sigma,\tau'}_{ex} = q'_{ex}= 1$.

\item $Type_{rand}$. Let $\sigma'$ be the strategy defined
by $\sigma'(\beta) = \sigma(\langle \epsilon, u \rangle \beta)$
for all $\beta \in V^*$.
Then, 
$$q^{k+1,\sigma,\tau'}_{u}  =   
\sum_{v} p_{u,v}  \: q^{k,\sigma',\tau'}_{v}
 \leq 
\sum_{v} p_{u,v} \:  q^{'}_{v} 
 = 
  q^{'}_{u}.$$

\item $Type_{call}$.   In this case, $u = (b,en) \in \Call_b$, and
$ q^{k+1,\sigma,\tau'}_{u} \leq 
\sup_{\rho}  q^{k,\rho,\tau'}_{en} \cdot \sup_{\rho}  q^{k,\rho,\tau'}_{(b,ex')} $,
where $ex' \in Ex_{Y(b)}$ is the unique exit node of $A_{Y(b)}$.
Now, by the inductive assumption, $q^{k,\rho,\tau'} \leq  q'$
for all $\rho$.
Moreover, since $q' = P(q')$, 
$q^{'}_{u} =
  q^{'}_{en} \cdot 
  q^{'}_{(b,ex')} $.
Hence, using these inequalities and substituting, we get
$$q^{k+1,\sigma,\tau'}_{u}
\leq  q^{'}_{en} \: 
    q^{'}_{(b,ex')} =  
    q^{'}_{u}.$$
\item $Type_{play}$: In this case,
starting at $\langle \epsilon,u \rangle$, whatever player 1's strategy $\sigma$ is,
it has the property that
$q^{k+1,\sigma,\tau'}_{u} \leq  \Val(A_u(q^{k,\sigma',\tau'}))$.
By the inductive hypothesis 
$q^{k,\sigma',\tau'}_{v} \leq   q^{'}_{v}$, so we are done by induction
and by the monotonicity of $\Val(A_u(x))$.\qed
\end{enumerate}

\noindent Now, by the lemma, $q^{*,\sigma,\tau'} =
\lim_{k \rightarrow \infty} \bfq^{k,\sigma,\tau'} \leq q'$.
This holds for any strategy $\sigma \in \Psi_1$.
Therefore,  $\sup_{\sigma \in \Psi_1} q^{*,\sigma,\tau'}_{u} \leq q'_{u}$,
for every vertex $u$.
Thus, by the determinacy of RCSG games, we have established that  
$q^*_{u} = \inf_{\tau \in \Psi_2} \sup_{\sigma \in \Psi_1} 
q^{*,\sigma,\tau}_{u}  \leq q'_{u}$, for all vertices $u$.
In other words, $q^* \leq q'$.
The fact that $\lim_{k \rightarrow \infty} P^k(\bfzero) \uparrow \bfqstar$
follows from a simple Tarski-Knaster argument.
\end{proof}

\begin{example}
{\rm 
For the system of equations $x=P(x)$ given in Example \ref{example:equations}, 
associated with the 1-RCSG given in  Example \ref{example:rcsg}, fairly easy calculations
using the equations show that  
the Least Fixed Point of the system (and thus the game values, starting
at the different vertices) is as follows:
$q^*_t  =  q^*_{(b_2,t)} =  1$;  $q^*_{u_5}  =  0$;
$q^*_s  = q^*_{u_1} = q^*_{u_2} = q^*_{u_4} = q^*_{(b_1,t)} = q^*_{(b_2,s)} =  0.5$;
$q^*_{u_3}  =  0.75$; and 
$q^*_{(b_1,s)}  =  0.25$.

In this case the values turn out to be rational and
are simple to compute, 
but in general the values may be irrational and
difficult to compute,  and
even if they are rational they may require exponentially many bits to represent
(in standard notation, e.g., via reduced numerator and denominator given in binary)
in terms of the size of the input 1-RCSG or equation system.

Furthermore, in this game there are pure optimal 
(stackless and memoryless) strategies for both
players.  Specifically,  the strategy for player 1 (maximizer) that always 
plays L from nodes $u_1$ is optimal, 
and the strategy for player 2 that always player L from
nodes $u_1$ and $u_4$ is optimal.
In general for 1-RCSGs,
we show randomized stackless and memoryless
$\epsilon$-optimal and optimal strategies do exist for players 1 and 2, 
respectively.  However,
for player 1 
only $\epsilon$-optimal strategies may exist, and
although optimal strategies do exist for player 2
they may require randomization using irrational probabilities. 
This is the case even for finite-state concurrent games. \qed }
\end{example}

We can use the system of equations to establish the following upper bound
for computing the value of a 1-RCSG termination game:

\begin{theorem}
The qualitative and quantitative termination problems
for 1-exit RCSGs can be solved in PSPACE. That is, 
given a 1-exit RCSG $A$, vertex $u$ and a rational probability $p$, 
there is a PSPACE algorithm to decide 
whether $\bfqstar_u \leq p$ (or $\bfqstar \geq p$, or $\bfqstar < p$, etc.).  
The running time is $O(|A|^{O(n)})$ 
where $n$
is the number of
variables in $\bfx = P(\bfx)$. 
We can also approximate the vector $\bfqstar$
of values  to within a specified number of
bits $i$ of precision ($i$ given in unary), in PSPACE and in time
$O(i |A|^{O(n)})$.
\label{thm:complexity}
\end{theorem}
\begin{proof}
Using the system $x=P(x)$,
we can express the condition $q^*_u \leq c$ 
by a sentence in the existential theory of the reals as follows:

$$ \exists x_1, \ldots, x_n  \bigwedge^n_{i = 1} 
(x_i = P_i(x_1,\ldots,x_n)) \wedge \bigwedge^n_{i = 1} 
(x_i \geq 0) \wedge (x_u \leq c)
$$

\noindent Note that the sentence is true, i.e. there exists a vector $x$ that
satisfies the constraints of the above sentence if and only if
the least fixed point $\bfqstar$ satisfies them.
The constraints $x_i = P_i(x_1,\ldots,x_n)$ for vertices $i$ of type
1, 2, and 3 (exit, probabilistic vertex and call port) are
clearly polynomial equations, as they should be in a sentence
of the existential theory of the reals.
We only need to show how to express equations of the
form $x_v = \Val(A_v(\bfx))$ in the existential theory of reals.
We can then appeal to well known results for deciding that theory
(\cite{Canny88,Ren92}). 
But this is a standard fact in game theory (see, e.g., 
\cite{BewKohl76,FiVr97,dAMa04} where it is used for finite CSGs).
The minimax theorem and its LP 
encoding allow the predicate ``$y = \Val(A_v(\bfx))$'' to be expressed
as an existential formula 
$\varphi(y,x)$ in the theory of reals with free variables $y$ and $x_1, \ldots, x_n$, such that for every $x \in \real^n$, there exists a unique $y$
(the game value) satisfying $\varphi(y,\bfx)$.
Specifically, the formula includes, besides the free variables $\bfx, y$,
existentially quantified variables $z_{\gamma_1}, \gamma_1 \in \Gamma^v_1$, and
$w_{\gamma_2}, \gamma_2 \in \Gamma^v_2$ for the probabilities of the
moves of the two players, and the conjunction of the following constraints
(recall that each entry $A_u(\gamma_1,\gamma_2)$ of the matrix $A_u$ is
a variable $x_v$ where $v$ is the vertex such that $(u,(\gamma_1,\gamma_2),v) \in \delta$) \smallskip
\begin{enumerate}[\qquad]
\item
$z_{\gamma_1} \geq 0$ for all $ \gamma_1 \in \Gamma^v_1$;
$~~~~~~\sum_{\gamma_1 \in \Gamma^v_1} z_{\gamma_1} =1$;
\item
$w_{\gamma_2} \geq 0$ for all $ \gamma_2 \in \Gamma^v_2$;
$~~~~~~\sum_{\gamma_2 \in \Gamma^v_2} w_{\gamma_2} =1$;
\item
$\sum_{\gamma_1 \in \Gamma^v_1} 
A_u(\gamma_1,\gamma_2) z_{\gamma_1} \geq y$ for all $ \gamma_2 \in
\Gamma^v_2$;
\item
$\sum_{\gamma_2 \in \Gamma^v_2} 
A_u(\gamma_1,\gamma_2) w_{\gamma_2} \leq y$ for all $ \gamma_1 \in
\Gamma^v_1$.
\end{enumerate}\smallskip

\noindent To approximate the vector of game values within given
precision we can do binary search using queries of the form $q^*_u
\leq c$ for all vertices $u$.
\end{proof} 
 
Determining the vertices $u$ for which the value $q^*_u$ is 0,
is easier and can be done in polynomial time, as in the case of
the turn-based 1-RSSGs \cite{EY06stacs}.

\begin{proposition} Given a 1-RCSG we can compute in polynomial time
the set $Z$ of vertices $u$ such that $q^*_u =0$. This set $Z$ depends only on
the structure of the given 1-RCSG and not on the actual values of the transition
probabilities.
\label{prop:zeros}
\end{proposition}
\begin{proof}
From the system of fixed point equations we have the following:
(1) all exit nodes are not in $Z$; (2) a probabilistic node $u$ is in $Z$
if and only if all its (immediate) successors $v$ are in $Z$;
(3) the call port $u=(b,en)$ of a box $b$ is in $Z$ if and only if
the entry node $en$ of the corresponding component $Y(b)$ is in $Z$
or the return port $(b,ex)$ is in $Z$;
(4) a play node $u$ is in $Z$ if and only if Player 2 has a move 
$\gamma_2 \in \Gamma^u_2$ such that for all moves 
$\gamma_1 \in \Gamma^u_1$ of Player 1, the next node $v$, i.e.
the (unique) node $v$ such that $(u,(\gamma_1,\gamma_2),v) \in \delta$,
is in $Z$.

Only the last case of a play node $u$ needs an explanation.
If Player 2 has such a move $\gamma_2$, then clearly the
corresponding column of the game matrix $A_u( \bfqstar)$ 
has all the entries 0, and the value of the game (i.e., $q^*_u$) is 0.
Conversely, if every column of $A_u( \bfqstar)$ has a nonzero entry,
then the value of the game with this matrix is positive
because for example Player 1 can give equal probability
to all his moves.
Thus, in effect, as far as computing the
vertices with zero value is concerned, 
we can fix the strategy of Player 1 at each
play vertex to play at all times all legal moves with equal
probability to get a 1-RMDP; a vertex has nonzero value
in the given 1-RCSG iff it has nonzero value in the 1-RMDP.

The algorithm to compute the set $Z$ of vertices
with 0 value is similar to the case of 1-RSSGs \cite{EY06stacs}.
Initialize $Z$ to $Q \setminus Ex$, the set of non-exit vertices.
Repeat the following until there is no change:
\begin{enumerate}[$\bullet$]
\item If there is a probabilistic node $u \in Z$ that has a 
successor
not in $Z$, then remove $u$ from $Z$.
\item If there is a call port $u=(b,en) \in Z$ such that
both the entry node $en$ of the corresponding component
$Y(b)$ and the return port $(b,ex)$ of the box are not in $Z$, 
then remove $u$ from $Z$.
\item 
If there is a play node $u \in Z$ such that for every move 
$\gamma_2 \in \Gamma^u_2$ of Player 2 there is a move 
$\gamma_1 \in \Gamma^u_1$ of Player 1 such that the next node $v$
from $u$ under $(\gamma_1,\gamma_2)$ is not in $Z$, 
then remove $u$ from $Z$.
\end{enumerate}

\noindent There are at most $n$ iterations and at the end $Z$ is the set
of vertices $u$ such that $q^*_u =0$. 
\end{proof}

\section{Strategy improvement and randomized-SM-determinacy}

\noindent The proof of Theorem 1 implies the following:

\begin{corollary}
\label{cor-min-player}
In every 1-RCSG termination game, player 2 (the minimizer) has an
optimal r-SM strategy.
\end{corollary}
\begin{proof}
Consider the strategy $\tau'$ in the proof of Theorem
\ref{lfp-char-thm},
chosen not for just any fixed point $\bfq'$, but for $\bfqstar$ itself.
That strategy is r-SM and is optimal.
\end{proof}

Player 1 does not have optimal
r-SM strategies, not even in finite concurrent stochastic games (see, e.g.,
\cite{FiVr97,dAMa04}).
We next establish that it does have finite r-SM  {\em $\epsilon$-optimal
strategies}, meaning that it has, for every $\epsilon > 0$, 
a r-SM strategy that guarantees
a value of at least $\bfq^*_u - \epsilon$, starting from every
vertex $u$ in the termination game.
We say that a game
is {\em r-SM-determined} if, 
letting $\Psi'_1$ and $\Psi'_2$ denote the set of r-SM 
strategies for players 1 and 2, respectively, we have
$\sup_{\sigma \in \Psi_1'} 
\inf_{\tau \in \Psi_2'} q^{*,\sigma,\tau}_{u}
= \inf_{\tau \in \Psi_2'}\sup_{\sigma \in \Psi_1'}  
q^{*,\sigma,\tau}_{u}$.

\begin{theorem} \mbox{}

\begin{enumerate}[\em(1)]
\item (Strategy Improvement) Starting at any r-SM strategy $\sigma_0$
  for player 1, via local strategy improvement steps at individual
  vertices, we can derive a series of r-SM strategies $\sigma_0,
  \sigma_1, \sigma_2, \ldots$, such that for all $\epsilon > 0$, there
  exists $i \geq 0$ such that for all $j \geq i$, $\sigma_j$ is an
  $\epsilon$-optimal strategy for player 1 starting at any vertex,
  i.e., $q^{*,\sigma_j}_u \geq q^*_u- \epsilon$ for all vertices $u$.

  \noindent Each strategy improvement step involves solving the quantitative
  termination problem for a corresponding 1-RMDP.  Thus, for classes
  where this problem is known to be in P-time (such as
  linearly-recursive 1-RMDPs, \cite{EY05icalp}), strategy improvement
  steps can be carried out in polynomial time.

\item Player 1 has $\epsilon$-optimal r-SM strategies, 
for all $\epsilon > 0$, in 1-RCSG termination games.

\item 1-RCSG termination games are r-SM-determined.

\end{enumerate}
\end{theorem}
\begin{proof}
Note that (2.) follows immediately from (1.),
and (3.) follows because 
by Corollary \ref{cor-min-player}, 
player 2 has an optimal r-SM strategy and thus \\
$\sup_{\sigma \in \Psi_1'} 
\inf_{\tau \in \Psi_2'} q^{*,\sigma,\tau}_{u}
= \inf_{\tau \in \Psi_2'}\sup_{\sigma \in \Psi_1'}  
q^{*,\sigma,\tau}_{u}$.

Let $\sigma$ be any r-SM strategy for player 1.
Consider $q^{*,\sigma}$. 
First, let us note that if $q^{*,\sigma} = P(q^{*,\sigma})$ 
then $q^{*,\sigma} = q^*$.  This is so because, by
Theorem \ref{lfp-char-thm}, $q^* \leq q^{*,\sigma}$,
and on the other hand, $\sigma$ is just one strategy for player 1, and 
for every vertex $u$,
$q^*_{u} =  
\sup_{\sigma' \in \Psi_1} 
\inf_{\tau \in \Psi_2} q^{*,\sigma',\tau}_{u}
\geq  \inf_{\tau \in \Psi_2} q^{*,\sigma,\tau}_{u}
= q^{*,\sigma}_{u}$.

Next we claim that, for all
vertices $u \not\in Type_{\play}$, 
$q^{*,\sigma}_{u}$  satisfies 
its equation in $x = P(x)$.  In other words,
$q^{*,\sigma}_{u} = P_{u}(q^{*,\sigma})$.
To see this, note that for vertices $u \not\in Type_{\play}$,
no choice of either player is involved, thus the equation
holds by definition of $q^{*,\sigma}$.
Thus, the only equations that may fail are those for $u \in Type_{play}$, 
of the form 
$x_{u} =  \Val(A_u(x))$.  We need the following.

\begin{lemma}
For any r-SM strategy $\sigma$ for player 1,  
and for any $u \in Type_{play}$, $q^{*,\sigma}_u \leq \Val(A_u(q^{*,\sigma}))$.
\end{lemma}
\begin{proof}
We are claiming that  $q^{*,\sigma}_u = \inf_{\tau \in \Psi_2} q^{*,\sigma,\tau}_u
\leq 
\Val(A_u(q^{*,\sigma}))$.
The inequality follows because a
strategy for player 2 can in the first step starting at vertex $u$ play
its optimal strategy in the matrix game $A_u(q^{*,\sigma})$,
and thereafter, depending on
which vertex $v$ is the immediate successor of $u$ in the play,
the strategy can play ``optimally'' to force at most the value
$q^{*,\sigma}_v$.  
\end{proof}

Now, suppose that for some $u \in Type_{play}$, 
 $q^{*,\sigma}_u \neq Val(A_u(q^{*,\sigma}))$.  Thus
by the lemma
$q^{*,\sigma}_u < Val(A_u(q^{*,\sigma}))$.
Consider a revised r-SM strategy for player 1,  $\sigma'$, which
is identical to $\sigma$, except that locally 
at vertex $u$ the strategy is changed so that $\sigma'(u) = p^{*,u,\sigma}$,
where $p^{*,u,\sigma} \in \mathcal{D}(\Gamma^u_1)$ is
an optimal mixed minimax strategy for player 1 in the
matrix game $A_u(q^{*,\sigma})$.
We will show that switching from 
$\sigma$ to 
$\sigma'$ will improve player 1's payoff at vertex
$u$, and will not reduce its payoff at any other vertex.

Consider a parameterized 1-RCSG, $A(t)$, which is identical to 
$A$, except that $u$ is a randomizing vertex, all edges
out of vertex $u$ are removed, and replaced by a single edge labeled
by probability variable $t$
to the exit of the same component, and an edge with remaining probability 
$1-t$
to a dead vertex.  
Fixing the value $t$ determines an 1-RCSG, $A(t)$.
Note that if we restrict the r-SM strategies $\sigma$ or 
$\sigma'$ to all vertices other
than $u$, then they both define the same r-SM strategy for the 1-RCSG $A(t)$.
For each vertex $z$ and strategy $\tau$ of player 2,
define $q^{*,\sigma,\tau,t}_{z}$ to be the probability of eventually terminating
starting from $\langle \epsilon,z \rangle$ 
in the Markov chain  $M^{z,\sigma,\tau}_{A(t)}$.
Let $f_z(t) =  \inf_{\tau \in \Psi_2} q^{*,\sigma,\tau,t}_{z}$.
Recall that $\sigma'(u) = 
p^{*,u,\sigma} \in \mathcal{D}(\Gamma^u_1)$ defines a probability
distribution on the actions available to player 1 at vertex $u$. 
Thus $p^{*,u,\sigma}(\gamma_1)$ is the probability of 
action $\gamma_1 \in \Gamma_1$.
Let $\gamma_2 \in \Gamma_2$ be any action of player 2
for the 1-step zero-sum game with game matrix $A_u(q^{*,\sigma})$.
Let $w(\gamma_1, \gamma_2)$ denote the vertex such that
$(u,(\gamma_1, \gamma_2), w(\gamma_1,\gamma_2)) \in \delta$.
Let $h_{\gamma_2} (t) = \sum_{\gamma_1 \in \Gamma_1} p^{*,u,\sigma}
(\gamma_1)   f_{w(\gamma_1,\gamma_2)} (t)$.

\begin{lemma}
\label{lem:pow-series}
Fix the vertex $u$.  Let $\varphi: \real \mapsto \real$ be  
any function $\varphi \in \{f_z \mid z \in Q\} \cup \{ h_\gamma \mid 
\gamma \in \Gamma^u_2\}$. 
The following
properties hold:
\begin{enumerate}[\em(1)]
\item If $\varphi (t) > t$ at some point $t \in [0,1]$, then 
$\varphi (t') > t'$ for
all $0 \leq t' <t$.
\item If $\varphi (t) < t$ at some point $t \in [0,1]$, then 
$\varphi (t') < t'$ for
all $1 > t' >t$.
\end{enumerate}
\end{lemma}
\begin{proof}

First, we prove this for $\varphi=f_z$, for some vertex $z$.

Note that, once player 1 picks a r-SM strategy, a 1-RCSG becomes a 1-RMDP.
By a result of \cite{EY05icalp}, player 2 has an optimal
deterministic SM response strategy. Furthermore, there is such a strategy that 
is optimal regardless of the starting vertex. 
Thus, for any value of $t$, player 2 has an optimal 
deterministic SM strategy  $\tau_{t}$,  such that for any start vertex $z$,  we have
$\tau_{t} = \arg \min_{\tau \in \Psi_2} q^{*,\sigma,\tau,t}_{z}$.
Let $g_{(z,\tau)}(t) = q^{*,\sigma,\tau,t}_{z}$, and 
let $d\Psi_2$ be the (finite) set of deterministic SM strategies of player 2. 
Then $f_z(t) = \min_{\tau \in d\Psi_2}  g_{z,\tau} (t)$.
Now, note that the function $g_{z,\tau}(t)$ is the probability of
reaching an exit in an RMC starting from a particular vertex. 
Thus, by \cite{EY04a}, $g_{z,\tau}(t) = (\lim_{k \rightarrow \infty} R^k(\bfzero))_{z}$
for a polynomial system $\bfx = R(\bfx)$ with non-negative coefficients, but with the
additional feature that the variable $t$ appears as one of the coefficients.
Since this limit can be described by a power series in the variable $t$
with non-negative coefficients, $g_{z,\tau}(t)$ has the following properties:
it is a continuous, differentiable, and 
non-decreasing function of $t \in [0,1]$,
with continuous and non-decreasing derivative,  $g'_{z,\tau}(t)$,
and since the limit defines probabilities we also know that
for $t \in [0,1]$, $g_{z,\tau}(t) \in [0,1]$.
Thus $g_{z,\tau}(0) \geq 0$ and $g_{z,\tau}(1) \leq 1$.

Hence, since  $g'_{z,\tau}(t)$ is non-decreasing, if for 
some $t \in [0,1]$, $g_{z,\tau}(t) > t$, 
then for all $t'<t$,  $g_{z,\tau}(t') > t'$.  To see this, note that if 
$g_{z,\tau}(t) > t$ and $g'_{z,\tau}(t) \geq 1$, then for all $t'' > t$, 
$g_{z,\tau}(t'')>t''$,
which contradicts the fact that $g_{z,\tau}(1) = 1$. 
Thus $g'_{z,\tau}(t) < 1$, and since $g'_{z,\tau}$ is non-decreasing,
it follows that  $g'_{z,\tau}(t') < 1$ for
all $t' \leq t$. Since $g_{z,\tau}(t) > t$,  we also have  $g_{z,\tau}(t') > t'$ 
for all $t' < t$.

Similarly, if $g_{z,\tau}(t) < t$ for some $t$, 
then  $g_{z,\tau}(t'') < t''$ for all $t'' \in [t,1)$.
To see this, note that if for some $t'' > t$, 
$t'' < 1$,  $g_{z,\tau}(t'') = t''$,
then since $g'_{z,\tau}$ is non-decreasing and 
$g_{z,\tau}(t) < t$, it must be the case that $g'_{z,\tau}(t'') > 1$.
But then $g_{z,\tau}(1)>1$, which is a contradiction.

It follows that $f_z(t)$ has the same properties, namely:
if  $f_z(t)>t$ at some point $t \in [0,1]$ then $g_{z,\tau}(t) >t$ for all $\tau$,
and hence for all $t' < t$ and for all $\tau \in d\Psi_2$, 
$g_{z,\tau}(t') > t'$, and thus 
$f_z(t') > t'$ for all $t' \in [0,t]$.
On the other hand, if $f_z(t)<t$  at $t \in [0,1]$, then 
there must be some $\tau'\in d\Psi_2$ such that  $g_{z,\tau'}(t) < t$. 
Hence $g_{z,\tau'}(t'') <t''$, for all $t'' \in [t,1)$, 
and hence $f_z(t'') < t''$ for all $t'' \in [t,1)$.

Next we prove the lemma for every $\varphi= h_{\gamma}$, where
$\gamma \in \Gamma^u_2$.
For every value of $t$, there is one SM strategy $\tau_t$  of
player 2 (depending  only on $t$)
that minimizes simultaneously $g_{z,\tau}(t)$ for all nodes $z$.
So $h_{\gamma} (t) = \min_{\tau} r_{\gamma,\tau}(t)$,
where $r_{\gamma,\tau}(t)= \sum_{\gamma_1 \in \Gamma_1} p^{*,u,\sigma}(\gamma_1)
g_{w(\gamma_1, \gamma),\tau} (t)$
is a convex combination (i.e., a ``weighted average'')
of some $g$ functions at the same point $t$.
The function $r_{\gamma,\tau}$ (for any subscript ) 
inherits the same properties as the $g$'s:
continuous, differentiable, non-decreasing, with continuous non-decreasing
derivatives, and 
$r_{\gamma,\tau}$ takes value between $0$ and $1$.
As we argued for the $g$ functions, in the same way it follows that 
$r_{\gamma,\tau}$ has properties 1 and 2.
Also, as we argued for $f$'s based on the $g$'s, it follows that $h$'s 
also have the same
properties, based on  the $r$'s.
\end{proof}

Now let $t_1 = q^{*,\sigma}_{u}$,
and let 
$t_2= \Val(A_u(q^{*,\sigma}))$.  By assumption, $t_2 > t_1$.
Observe that
$f_z(t_1) = q^{*,\sigma}_{z}$ for every vertex $z$. 
Thus, 
$h_{\gamma_2} (t_1) = \sum_{\gamma_1 \in \Gamma_1} p^{*,u,\sigma}(\gamma_1)  
f_{w(\gamma_1,\gamma_2)} (t_1)  = 
\sum_{\gamma_1} p^{*,u,\sigma}(\gamma_1) 
q^{*,\sigma}_{w(\gamma_1,\gamma_2)}$.
But since, by definition, $p^{*,u,\sigma}$ is an optimal strategy
for player 1 in the matrix game $A_u(q^{*,\sigma})$, it must
be the case that for every $\gamma_2 \in \Gamma^u_2$, 
$h_{\gamma_2}(t_1) \geq t_2$, for otherwise
player 2 could play a strategy against $p^{*,u,\sigma}$ which
would force a payoff lower than the value of the game. 
Thus $h_{\gamma_2}(t_1) \geq t_2 > t_1$, for all $\gamma_2$.
This implies that $h_{\gamma_2}(t)>t$ for all $t<t_1$ by Lemma 2, 
and for all $t_1 \leq t<t_2$, because 
$h_{\gamma_2}$ is non-decreasing. 
Thus, $h_{\gamma_2}(t)>t$ for all $t<t_2$.

Let $t_3 = q_u^{*,\sigma'}$.
Let $\tau'$ be an optimal global 
strategy for player 2 against $\sigma'$; 
by \cite{EY05icalp}, we may assume $\tau'$ is a deterministic SM strategy.
Let $\gamma'$ be player 2's action in $\tau'$ at node $u$.
Then the value of any node $z$ under the pair of strategies $\sigma'$ and
$\tau'$ is 
$f_z(t_3)$, and thus since $h_{\gamma'} (t_3)$ is
a weighted average of $f_z(t_3)$'s for some set of $z$'s, 
we have $h_{\gamma'}(t_3)  = t_3$.
Thus, by the previous paragraph, it must be that $t_3 \geq t_2$, and
we know $t_2 > t_1$.
Thus, $t_3 = q^{*,\sigma'}_u \geq \Val(A_u(q^{*,\sigma})) > t_1 = 
q^{*,\sigma}_u$. 
We have shown:

\vspace{-0.05in}

\begin{lemma}
\label{lem:improv-to-val}
$q^{*,\sigma'}_u \geq  \Val(A_u(q^{*,\sigma})) > q^{*,\sigma}_u$.
\end{lemma}

\vspace{-0.05in}

Note that since $t_3 > t_1$, and $f_z$ is 
non-decreasing, we have $f_z(t_3) \geq f_z(t_1)$ 
for all vertices $z$.  But then 
$q^{*,\sigma'}_{z} =f_z(t_3) \geq f_z(t_1) = q^{*,\sigma}_{z}$  
for all $z$.
Thus, $q^{*,\sigma'} \geq q^{*,\sigma}$, with strict inequality
at $u$, i.e., $q^{*,\sigma'}_u > q^{*,\sigma}_u$.
Thus, we have established that such a ``strategy improvement''
step does yield a strictly better payoff for player 1.

Suppose we conduct this ``strategy improvement'' step repeatedly,
starting at an arbitrary initial r-SM strategy  $\sigma_0$, as long
as we can. This leads to a  (possibly infinite) sequence of r-SM strategies 
$\sigma_0, \sigma_1, \sigma_2, \ldots$.
Suppose moreover, that during these improvement
steps  we always ``prioritize'' among vertices at which to improve so that,
among all those vertices $u \in Type_{play}$ which can be improved,
i.e.,
such that $q^{*,\sigma_i}_u < \Val(A_u(q^{*,\sigma_i}))$, we choose
the vertex which has not been improved for the longest number of
steps (or one that has never been improved yet).  This insures that, 
infinitely often,
at every vertex at which the local strategy can be improved, it 
eventually is improved.

Under this strategy improvement regime, we show that $\lim_{i \rightarrow \infty} 
q^{*,\sigma_i} = q^*$, and thus, for 
all $\epsilon > 0$, there exists a sufficiently
large  $i \geq 0$ such that $\sigma_i$ is an  $\epsilon$-optimal
r-SM strategy for player 1.
Note that after every strategy improvement step, $i$, which 
improves at a vertex $u$,
by Lemma \ref{lem:improv-to-val} 
we will have $q^{*,\sigma_{i+1}}_u \geq \Val(A_u(q^{*,\sigma_i}))$.
Since our prioritization assures
that every vertex that can be improved at any step $i$ will be improved eventually, for all $i \geq 0$ there exists $k \geq 0$ such that
$q^{*,\sigma_i} \leq P(q^{*,\sigma_i}) \leq q^{*,\sigma_{i+k}}$.
In fact, there is a uniform bound on $k$, namely $k \leq |Q|$,  the number of vertices.
This ``sandwiching'' property allows us to conclude that, in the 
limit, this sequence reaches a fixed point of $x=P(x)$.
Note that since $q^{*,\sigma_i} \leq q^{*,\sigma_{i+1}}$  for all
$i$, and since $q^{*,\sigma_i} \leq q^*$, we know
that the limit $\lim_{i \rightarrow \infty} q^{*,\sigma_i}$
exists.  Letting this limit be $q'$, we have $q' \leq q^*$.
Finally, we have $q' = P(q')$, because 
letting $i$ go to infinity 
in all three parts of the ``sandwiching'' inequalities 
above, we get 
$q' \leq \lim_{i \rightarrow \infty} P(q^{*,\sigma_i}) \leq q'$.
But note that $\lim_{i \rightarrow \infty}
P(q^{*,\sigma_i}) = P(q')$, because the mapping $P(x)$ is continuous
on $\real^n_{\geq 0}$.
Thus $q'$ is a fixed point of $x = P(x)$, and $q' \leq q^*$.
But since $q^*$ is the {\em least} fixed point of $x= P(x)$, we 
have $q' = q^*$.  
\end{proof}

We have so far not addressed the complexity of computing or
approximating the ($\epsilon$-) optimal strategies for the two players
in 1-RCSG termination games.  Of course, in general, player 1
(maximizer) need not have any optimal strategies, so it only makes
sense to speak about computing $\epsilon$-optimal strategies for it.
Moreover, the optimal strategies for player 2 may require
randomization that is given by irrational probability distributions
over moves, and thus we can not compute them exactly, so again we must
be content to approximate them or answer decision questions about
them.  It is not hard to see however, by examining the proofs of our
theorems, that such decision questions can be answered using queries
to the existential theory of reals, and are thus also in PSPACE.

\section{Lower bounds}

\noindent Recall the {\em square-root sum problem} (e.g., from
\cite{GGJ76,EY04a}): given $(a_1,\ldots,a_n) \in \nat^n$ and $k \in
\nat$, decide whether $\sum^n_{i=1} \sqrt{a_i} \geq k$.

\begin{theorem}
\label{thm:sqrt-csg-reduction}
There is a P-time reduction from the square-root sum problem
to the quantitative termination (decision) problem for finite CSGs.
\end{theorem}

\begin{proof}
Given positive integers $(a_1,\ldots,a_n) \in \nat^n$, and $k \in \nat$,
we would like to check whether $\sum^n_{i=1} \sqrt{a_i} \geq k$.
We can clearly assume that $a_i > 1$ for all $i$.
We will reduce this problem to the problem of deciding whether 
for a given finite CSG, starting at a given node, the value of the
termination game is greater than a given rational value.

Given a positive integer $a > 1$, we 
will construct  a finite CSG, call it gadget $G(a)$, 
with the property that for a certain node $u$ in $G(a)$ the value of the 
termination game starting at $u$  is
$d+e \sqrt{a}$,
where $d$ and $e$ are  
rationals that depend on $a$, with $e > 0$,
and such that we can compute
$d$ and $e$ efficiently, in polynomial time, given $a$.

If we can construct such gadgets, then we can do the reduction as follows.
Given $(a_1, \ldots, a_n) \in \nat^n$, with $a_i > 1$ for all 
$i$, 
and given $k \in \nat$,
make copies of the gadgets $G(a_1)$, \ldots, $G(a_n)$. 
In each gadget $G(a_i)$ we have a node $u_i$ whose termination value 
is $d_i + e_i \sqrt{a_i}$,  where $d_i$ and $e_i > 0$ 
are rationals that depend on $a_i$ and can 
be computed efficiently from $a_i$.
Create a new node $s$ and add
transitions from $s$ to the nodes $u_i$, $i= 1 \ldots,n$, with probabilities 
$p_i = E/ e_i$, respectively,  where $E = 1/(\sum^n_{i=1} \frac{1}{e_i})$.
It is easy to check that the value of termination starting at $s$ is 
$D+E \sum^n_{i=1} \sqrt{a_i}$,
where $D = \sum^n_{i=1} p_i d_i$. 
Note that $D$ and $E$ are rational values that we can compute efficiently
given the $a_i$'s,
so to solve the square root sum problem, i.e.,
decide whether $\sum^n_{i=1} \sqrt{a_i} \geq k$, 
we can ask whether the
value of the termination game starting at node $s$ is  $\geq D + E k$.

Now we show how to construct the gadget $G(a)$ given a positive integer $a$.
$G(a)$ has a play node $u$, the target node $t$, dead node $z$,
and probabilistic nodes $v_1$, $v_2$. Nodes $z$ and $t$ are absorbing.
At $u$ each player has two moves $\{1,2\}$.
If they play $1,1$ then $u$ goes to $v_1$,
if they play $2,2$ then $u$ goes to $v_2$,
if they play $1,2$ or $2,1$ then $u$ goes to $z$.

Note that we can write $a$ as $a= m^2-l$ where $m$ is a small-size rational
($m$ is approximately $\sqrt{a}$)
and $l < 1$ is also a small-size rational, and such that we can compute 
both $m$ and $l$ efficiently given $a$.
To see this  note that,  first, 
given $a$ we can easily approximate $\sqrt{a}$ from above
to within an additive error at most $1/(2a)$ in polynomial time,
using standard methods for approximating square roots. 
In other words, given integer $a >1$,
we can efficiently compute a rational number $m$ 
such that $0 \leq m - \sqrt{a} \leq 1/(2a)$.
We then have
\begin{eqnarray*}
               m^2 & \leq & (\sqrt{a} + 1/(2a))^2\\
             & = & a + 1/\sqrt{a} + 1/(4 a^2)
\end{eqnarray*}
Since  $1/ \sqrt{a}  + 1/ (4 a^2) < 1$,
we can let $l =  m^2 - a$.

Having computed $m$ and $l$, 
let $c_2=l/4$, $g=m-1-c_2$, and $c_1 = g c_3$, where $0< c_3 < 1$ 
is a  small-sized rational value such that $c_3 <  1/2g$.
From node $v_1$ we move with probability $c_1$ to $t$, with probability $c_2$ 
to $u$, and
with the remaining probability to $z$.
From node $v_2$ we go with probability $c_3$ to $t$ and $1-c_3$ to $z$.
It is not hard to check that these are
legitimate probabilities.

Let $x$ be the value at $u$.
We have $x = \Val(A)$, where
the $2 \times 2$ matrix $A$ for the one-shot zero-sum matrix game at $u$ has 
$A_{1,1}= c_1+ c_2 x$,
$A_{2,2}= c_3$, and $A_{1,2}=A_{2,1}=0$.  
Note that $A_{1,1} > 0$ and $A_{2,2} > 0$.
If the optimal strategy of player 1 at $u$ is to play $1$ with probability $p$
and $2$ with probability $1-p$, then 
by basic facts about zero-sum matrix games we must have $0 < p < 1$ and
$x = p (c_1+c_2 x) = (1-p)c_3$.
So $p = c_3/ (c_1+c_2 x+c_3)$, and substituting this expression for $p$ in 
the equality $x= p (c_1 + c_2 x)$, 
we have:
\[ c_2 x^2 + (g c_3+c_3 - c_2c_3)x -g(c_3)^2 = 0 \]
So, \[x = \frac{ -(gc_3+c_3 - c_2c_3) + \sqrt{ (gc_3+c_3 - c_2c_3)^2 + 4g c_2(c_3)^2 }} {2c_2}\]
Note that we must choose the root with $+$ sign to get a positive value.

The discriminant can be written as $(c_3)^2 [ (g+1-c_2)^2 + 4g c_2]$.
The term $(c_3)^2$  will come out from under the square root, as $c_3$, so we care only about the 
expression in
the brackets, which is 
\begin{eqnarray*}
(g+1-c_2)^2 + 4g c_2
&= &(g+1)^2 + (c_2)^2 - 2gc_2 -2c_2 +4gc_2\\
&= & (g+1)^2 + (c_2)^2 + 2gc_2 +2c_2 -4c_2\\ & = & (g+1+c_2)^2 - 4c_2\\
&=&  m^2 -l\\ & = &a
\end{eqnarray*}
So $x = d + e \sqrt{a}$, where $d = -(gc_3+c_3 - c_2c_3)/2c_2$
and $e = c_3/2c_2$.
\end{proof}

 \begin{theorem} 
 \label{thm:csg-reduction}
 There is a P-time reduction  
 from the quantitative termination (decision) problem for finite CSGs to the
 qualitative termination problem for 1-RCSGs.
 \end{theorem}

\begin{proof}
Consider the 1-RMC depicted in Figure \ref{fig:example_rmc_param}. 
We assume $p_1+ p_2 = 1$.  As shown in (\cite{EY04a}, Theorem 3),
in this
 1-RMC
the probability of termination starting at $\langle \epsilon, en \rangle$
is $= 1$ if and only if $p_2 \geq 1/2$.

Now, given a finite CSG, $G$, and a vertex $u$ of $G$, do the following:
first ``clean up'' $G$ by removing all
nodes where the min player (player 2) has a strategy to achieve probability 0.
We can do this in polynomial time as follows.
Note that the only way player 2 can force a probability 0 of termination 
is if it has a strategy $\tau$ such that, for all strategies $\sigma$ of
player 1, there is no path in the resulting Markov chain from the start
vertex $u$ to the terminal node.  But this can only happen if, ignoring
probabilities, player 2 can play in such a way as to avoid the terminal vertex.
This can be checked easily in polynomial time.

\begin{figure*}[t]
\begin{center}
\scalebox{0.39}{\input{example_param_rmc_exactly1.pstex_t}}
\caption{1-RMC $A'$}
\label{fig:example_rmc_param}
\end{center}
\end{figure*}

The revised CSG will have two designated terminal nodes, the old 
terminal node, labeled ``1'', and another terminal node labeled ``0''.
From every node $v$  of $Type_{rand}$ in the revised CSG which does not carry full probability on
its outedges, we direct all the ``residual'' probability to ``0'', i.e.,
we add an edge from $v$ to ``0'' with probability
$p_{v,\mbox{``0''}} = 1 - \sum_{w} p_{v,w}$, where the sum is over
all remaining nodes $w$ is the CSG.

Let $\epsilon > 0$ be a value that is strictly less than the least probability,
over all vertices,
under any strategy for player 2, of reaching the terminal node.
Obviously such an $\epsilon > 0$ exists in the revised CSG, because by 
Corollary \ref{cor-min-player} (specialized to the case of finite CSGs) 
player 2 has an optimal randomized S\&M strategy.
Fixing that strategy $\tau$, player 1 can force termination from
vertex $u$ with 
positive probability $q^{*,\cdot,\tau}_u$.  We take 
$\epsilon =  (\min_u q^{*,\cdot,\tau}_u) / 2$. (We do not need
to compute $\epsilon$; 
we only need its existence for the correctness proof of the reduction.)

In the resulting finite CSG, we know that if  player 1
plays $\epsilon$-optimally (which it can do with randomized S\&M 
strategies), and player 2 plays arbitrarily,
there is no bottom SCC in the resulting finite Markov
chain other than the two designated terminating nodes ``0'' and ``1''.
In other words, all the probability exits the system, as
long as the maximizing player plays $\epsilon$-optimally.

Now, take the remaining finite CSG, call it $G'$.
Just put a copy of $G'$ at the entry of the component $A_1$ 
of the 1-RMC
in Figure \ref{fig:example_rmc_param},
identifying the entry $en$ with the initial node, $u$, of $G'$.
Take every transition that is directed into the terminal node ``1'' of G,
and instead direct it to the exit $ex$ of the component $A_1$.
Next, take every edge that is directed into the terminal ``0'' node
and direct it to the first call port, $(b_1,en)$  of the left box $b_1$.
Both boxes map to the unique component $A_1$.
Call this 1-RCSG $A$.

We now claim that the value $q^*_{u} \geq 1/2$ in the finite CSG $G'$ for
terminating at the terminal ``1'' iff the value $q^*_u =1$  
for terminating in the resulting
1-RCSG, $A$.
The reason is clear: after cleaning up the CSG, we
know that under an $\epsilon$-optimal strategy
for the maximizer for reaching ``1'', all the probability exits $G'$ either at 
``1'' or at ``0''.
We also know that the supremum value that the maximizing player 
can attain will have value 1 iff 
the supremum probability it can attain  for going directly to the 
exit of the component in $A$ is $\geq 1/2$, but this is precisely the
supremum probability that maximizer can attain for going to ``1'' in $G'$.

Lastly, note that the fact that the quantitative probability was
taken to be $1/2$ for the finite CSG is without loss of generality.
Given a finite CSG $G$ and a rational probability $p$, $0 < p < 1$,  
it is easy to efficiently construct another finite CSG $G'$
such that the termination probability for $G$ is $\geq p$ iff
the termination probability for $G'$ is $\geq 1/2$.  
\end{proof}

\section{Conclusions}

\noindent We have studied Recursive Concurrent Stochastic Games
(RCSGs), and we have shown that for 1-exit RCSGs with the termination
objective we can decide both quantitative and qualitative problems
associated with computing their values in PSPACE, using decision
procedures for the existential theory of reals, whereas any
substantial improvement (even to NP) of this complexity, even for
their qualitative problem, would resolve a long standing open problem
in exact numerical computation, namely the square-root sum problem.
Furthermore, we have shown that the quantitative decision problem for
finite-state concurrent stochastic games is also at least as hard as
the square-root sum problem.

An important open question is whether approximation of the game values, to
within a desired additive error $\epsilon > 0$, for both
finite-state concurrent games and for 1-RCSGs, can be done more efficiently.
Our lower bounds (with respect to square-root sum)  do not address the approximation question, and
it still remains open whether (a suitably formulated
gap decision problem associated with) approximating the value of even 
finite-state  CSGs, 
to within a given additive error $\epsilon > 0$, is in NP.

In \cite{EY05icalp}, we showed that model checking linear-time 
($\omega$-regular
or LTL) properties for 1-RMDPs (and thus also for 1-RSSGs)
is undecidable, and that even the qualitative or approximate
versions of such linear-time model checking questions remains undecidable.  
Specifically,  for any $\epsilon > 0$, 
given as input a 1-RMDP and an LTL property, $\varphi$, 
it is undecidable to determine whether 
the optimal probability with which the controller can force 
(using its strategy) the executions
of the 1-RMDP to satisfy $\varphi$, is probability $1$, or is at most
probability $\epsilon$, 
even when
we are guaranteed that the input satisfies one of these two cases.
Of course these undecidability results extend to the more general 1-RCSGs.

On the other hand, building on our polynomial time algorithms for the qualitative
termination problem for 1-RMDPs 
in \cite{EY06stacs},  Br\'{a}zdil et. al. \cite{BBFK06}
showed decidability (in P-time) for the qualitative problem of deciding whether 
there exists a strategy under which a given target vertex 
(which may not be an exit) of a 1-RMDP 
is reached in {\em any} calling context (i.e., under any call stack)  
almost surely (i.e., with probability 1).
They then used this decidability result to show that the qualititive
model checking problem
for 1-RMDPs against a qualitative fragment of the branching time
probabilistic temporal logic PCTL is decidable.

In the setting of 1-RCSGs (and even 1-RSSGs), it remains an open problem whether
the qualitative problem of reachability of a vertex (in any calling context)
is decidable.  Moreover, it should be noted that even for 1-RMDPs,
the problem of deciding whether the {\em value} of the reachability game is 1
is not known to be decidable.  This is because although the result
of \cite{BBFK06} shows that it is decidable whether there exists a strategy
that achieves probability 1 for reaching a desired vertex, there may
not exist any optimal strategy for this reachability problem, 
in other words the value may be 1 but it may only be attained as
the supremum value achieved over all strategies.

 \noindent {\bf Acknowledgement}  We thank Krishnendu Chatterjee
 for helpful discussions clarifying several results about finite CSGs obtained 
 by himself and others.  This work was partially supported by NSF grants CCF-04-30946
and CCF-0728736.

\bibliographystyle{plain}

\end{document}